\documentclass[oldversion]{aa}
\usepackage{txfonts}
\usepackage{graphicx}
\usepackage[dvips]{color}
\usepackage{natbib}
\usepackage{supertabular}
\usepackage{color}

\bibpunct{(}{)}{;}{a}{}{,}
\begin{document}

\title{Gravitational lenses as cosmic rulers: $\Omega_{\rm m}$, $\Omega_{\Lambda}$ from time delays and velocity dispersions}

\author{Danuta Paraficz \inst{1}, Jens Hjorth  \inst{1}}
\institute{Dark Cosmology Centre, Niels Bohr Institute, University of Copenhagen,
		Juliane Maries Vej 30, DK-2100 Copenhagen, Denmark}

\date{\today}
\authorrunning{Danuta Paraficz} 
\titlerunning{Gravitational lenses as cosmic rulers: $\Omega_{\rm m}$, $\Omega_{\Lambda}$ from time delays and velocity dispersions}

\abstract{We show that a cosmic standard ruler can be constructed from the joint measurement
of the time delay, $\Delta \tau$, between gravitationally lensed quasar images and the 
velocity dispersion, $\sigma$, of the lensing galaxy. This is specifically shown, for a
singular isothermal sphere lens,
$D_{OL} \propto  \Delta \tau / \sigma^2$, where $D_{OL}$ is the angular diameter 
distance to the lens. Using MCMC simulations
we illustrate the constraints set in the $\Omega_{\rm m}$--$\Omega_{\Lambda}$ plane
from future observations.}

\keywords{gravitational lensing, cosmological parameters}

\maketitle

\section{Introduction}
In the cosmological $\Lambda$CDM model, it is currently estimated that 96\% of the total energy density of the Universe 
is in the form of dark matter (25\%) and dark energy (74\%). These proportions
have been inferred from a number of completely independent measurements, e.g. cosmic
microwave background (CMB)  \citep{Spergel:1999,Komatsu:2009}, baryon acoustic oscillations \citep{Eisenstein:2005},
supernovae  \citep{Riess:1998,Perlmutter:1999}, large scale structure  \citep{Peacock:2001}, clusters of galaxies \citep{White:1993} and weak gravitational
lensing \citep{Weinberg:2003}. Unfortunately,  any astrophysical approach suffers
from potential systematic uncertainties   (e.g. supernovae -- not standard but standardized candles with possible redshift evolution;  CMB -- various parameter degeneracies and interference with foregrounds; weak lensing -- PSF influencing the galaxy shape measurement). It is therefore important to explore complementary
methods for measuring these quantities.
Gravitationally lensed quasars QSOs offer such an attractive alternative.

QSOs that are positioned in a way that a lens, i.e., a massive
foreground object such as a galaxy or a group of galaxies, intersects the line of sight, can be seen as magnified, 
  multiple images. Gravitationally lensed QSOs have already been used to set constraints 
on cosmological parameters. Notably
\citet{Refsdal:1964} showed that the Hubble constant $H_0$ can be measured from a 
multiple imaged QSO if the time delay between the lensed images and the mass distribution of the lens are known;   
attempts to constrain the cosmological constant $\Lambda$
have been based on gravitationally lensed QSO statistics \citep{Fukugita:1990}. 
The importance of strong gravitational lensing in future constraints 
on the evolution of the dark energy equation of state parameter $w(z)$, has
also been emphasized \citep{Linder:2004}.

In recent years, joint studies of stellar dynamics and gravitational lensing have proven
very fruitful, e.g. the Lenses Structure \& Dynamics (LSD) Survey and the Sloan Lens ACS Survey (SLACS) were used to constrain the 
density profiles of galaxies \citep{Treu:2004,Koopmans:2006b}.  
Methods for using either velocity dispersions \citep{Grillo:2008} or time delays 
\citep{Dobke:2009} of lensed systems have been proposed as estimators of 
$\Omega_{\rm m}$ and $\Omega_\Lambda$.

We note that a standard cosmic ruler can be constructed
from the joint measurement of the time delay between QSO images and the velocity 
dispersion of the lensing galaxy, independent of the redshift of the QSO. We
explore its use to constrain $\Omega_{\rm m}$ and $\Omega_\Lambda$ in view
of the large samples of lenses to be found in forthcoming experiments, such as from the Large Synoptic Survey Telescope (LSST),  Square Kilometre Array (SKA), Joint Dark Energy Mission ({\em JDEM}), {\em Euclid}, and the Observatory for Multi-Epoch Gravitational Lens Astrophysics ({\em OMEGA}) \citep{Dobke:2009,Ivezic:2007,Carilli:2004,Marshall:2005,Moustakas:2008}.

\section{Lenses as standard rulers}

The time delay $ \Delta\tau$ is the combined effect of 
the difference in length of the
optical path between two images and the gravitational time dilation of
two light rays passing through different parts of the lens potential well,
\begin{equation}
  \Delta\tau = \frac{1+z_{\rm L}}{c}\frac{D_{\rm OS}D_{\rm OL}}{D_{\rm LS}}
  \left( \frac{1}{2} (\vec{\theta}-\vec{\beta})^2-\Psi(\vec{\theta})\right).
  \end{equation}
Here $\vec{\theta}$ and $\vec{\beta}$ are the positions of the images and the source respectively, $z_{\rm L}$ is the lens redshift, and $\Psi$ is the effective gravitational potential of the lens.  
$D_{\rm OL}$, $D_{\rm OS}$, $D_{\rm LS}$ are the angular diameter distances
between observer and lens, observer and source, and lens and
source, respectively. 
If the lens geometry $\vec{\theta} - \vec{\beta}$ and
the lens potential $\Psi$ are known, the time delay measures the
ratio $D_{\rm OS}D_{\rm OL}/D_{\rm LS}$, also known as the
effective lensing distance $r(z,\Omega_{\rm m},\Omega_\Lambda)$,
which depends on the cosmological parameters.

The observed velocity dispersion of a galaxy $\sigma$ is the result 
of the superposition of
many individual stellar spectra, each of which has been Doppler shifted
because of the random stellar motions within the galaxy. Therefore, it can be determined by
analyzing the integrated spectrum of the galaxy,  which has 
broadened absorption lines due to the motion of the stars.
The velocity dispersion is related to the mass through the virial theorem: $\sigma^2\propto M_{\sigma}/R$, where $M_{\sigma}$ is the mass enclosed inside the radius $R$.
The  mass is measured by the Einstein angle $\theta_{\rm E}$ of the lensing system $M_{\theta_{\rm E}}=\frac{c^2}{4G}  \frac{D_{\rm OL}D_{\rm OS}}{D_{\rm LS}}\theta_{\rm E}^2$, where $R=D_{\rm OL}\theta_{\rm E}$. Thus,
$\sigma^2 \propto \frac{D_{\rm OS}}{D_{\rm LS}}\theta_{\rm E}$.

Since the time delay is proportional to
$D_{\rm OS}D_{\rm OL}/D_{\rm LS}$ and the velocity dispersion is proportional to $D_{\rm OS}/D_{\rm LS}$, the ratio $\Delta \tau/\sigma^2$  is dependent only on the lens distance and therefore acts as a cosmic ruler: 
$\Delta\tau/\sigma^2 \propto
D_{\rm OL}$ \footnote{A standard ruler can be used to measure angular diameter distances. Standard candles, on the other hand, measure luminosity distances, which can be obtained by
multiplying angular diameter distances by a factor $(1+z)^2$.}.
In Fig.~1 we illustrate  the dependency on the lens redshift of the
measurables,  time delay $\Delta \tau$,
velocity dispersion squared $\sigma^2$, and the ratio between them,
$\Delta\tau/\sigma^2$. To show the sensitivity of the three functions to $\Omega_\Lambda$, we plot them for four cases of a flat ($\Omega_{\rm m}+\Omega_{\Lambda}=1$) Universe with $\Omega_{\Lambda}=$  0.2, 0.5, 0.7 and 0.9, relative to an Einstein--de Sitter Universe  
($\Omega_{\rm m}=1, \Omega_{\Lambda}=0$).

We can see that $\Delta \tau/\sigma^2$  is more sensitive to the cosmological parameters
than $\Delta \tau$ or $\sigma^2$ separately. We also note that the higher the lens redshift, the more pronounced is the dependency on cosmology,  hence it is important  for this method to study high redshift  lenses. Finally, it has the advantage of being independent of the source redshift.
\begin{center}
\begin{figure}[!htbp]
\includegraphics[scale=0.5]{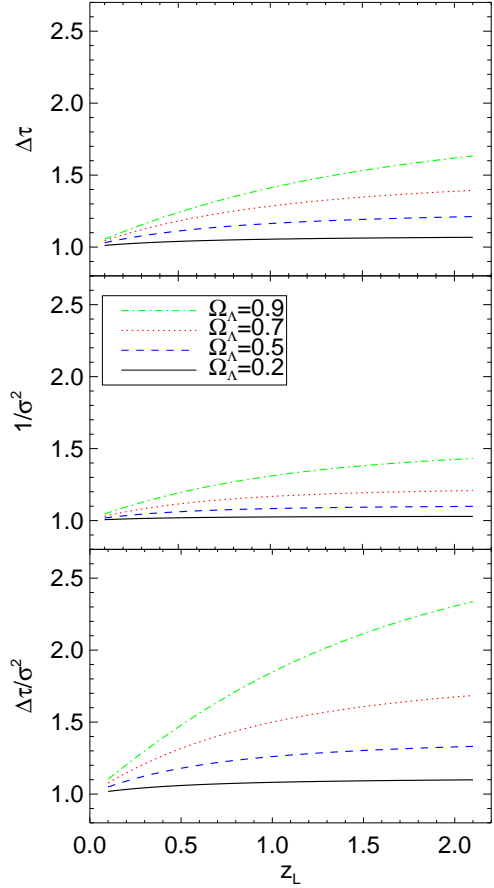}
\caption{Dependence of the three quantities, $\Delta\tau$ ({\it top panel}), $\sigma^2$ ({\it middle panel}), $\Delta\tau/\sigma^2$ ({\it bottom panel}) on the lens redshift. The source redshift was fixed to $z=3$. A flat Universe ($\Omega_{\rm m}+\Omega_{\Lambda}=1$)  was assumed with 4 different values $\Omega_{\Lambda}$: 0.2, 0.5, 0.7 and 0.9. Each curve is plotted  relative to a flat Einstein--de Sitter Universe ($\Omega_{\rm m}=1$, $\Omega_{\Lambda}=0$).}
\end{figure}
\end{center}

Both velocity dispersion and time delay depend on the gravitational potential of the lens galaxy. For the simple case of a singular isothermal sphere (SIS)
both parameters can be easily expressed analytically. While the SIS model is convenient for its simplicity, it is also 
a surprisingly useful model for lens galaxies 
\citep{Koopmans:2006b,Oguri:2007,Schechter:2004,Guimaraes:2009,Koopmans:2009}. 

The time delay is
\begin{equation}
  \Delta\tau_{\rm SIS} = \frac{1+z_{\rm L}}{2c}
  \frac{D_{\rm OL}D_{\rm OS}}{D_{\rm LS}}(\theta_2^2-\theta_1^2)
  \end{equation}
and the velocity dispersion  is
\begin{equation}
 \sigma^2_{\rm SIS} = \theta_{\rm E}\frac{c^2}{4\pi}\frac{D_{\rm OS}}{D_{\rm LS}}.
\end{equation}
For the SIS model the Einstein angle is  half the distance between the lensed images, $\theta_{\rm E}=(\theta_1+\theta_2)/2$ (we define
$\Delta\tau$ and $\theta_{1,2}$ to be positive, with $\theta_2 \ge \theta_1$).
Hence 
\begin{equation}
D_{\rm OL} (\theta_2-\theta_1) = \frac{c^3}{4\pi} \frac{\Delta\tau_{\rm SIS}} {\sigma^2_{\rm SIS}(1+z_{\rm L}) }. 
\end{equation}

\section{Monte Carlo Markov Chain simulations}

To explore the constraints set on cosmological parameters from the joint
measurements of image positions, lens redshift, time delay and velocity dispersion, we perform
simulations on SIS lenses using the Metropolis algorithm \citep{Saha:1994}. 
To simulate the uncertainties of the measurables we use $\chi^2$ given by:

\begin{equation}
\chi^2=  \frac{
  \left[\frac{\Delta\tau}{\sigma^2\theta_{\rm E}(1+z_{\rm L})}
    \frac{c^3}{8\pi}  - r(z,\Omega_{\rm
    m},\Omega_\Lambda)\right]^2}
  {\left(\frac{c^3}{8\pi}\right)^2\left[
  \left(\frac{\Delta\tau}{\sigma^2\theta_{\rm E}^2}\frac{\delta\theta_{\rm E}}{1+z_{\rm L}}\right)^2 +
  \left(\frac{\Delta\tau}{\sigma^4\theta_{\rm E}}\frac{\delta\sigma^2}{1+z_{\rm L}} \right)^2+
  \left(\frac{\delta\Delta\tau}{\sigma^2\theta_{\rm E}(1+z_{\rm L})}\right)^2
  \right]},
\end{equation}
 where, for simplicity  we have assumed that for each simulated lens, the lens images are aligned such that $\theta_1=0$ and $\theta_2=2\theta_{\rm E}$.
Similar expressions are computed for $\Delta\tau$ and $\sigma^2$. In all the simulations we let $\Omega_{\rm m},\Omega_\Lambda$ vary between 0 to 1 while either fixing or marginalizing over $H_0$.
\begin{figure}[!htbp]
\includegraphics[scale=0.5]{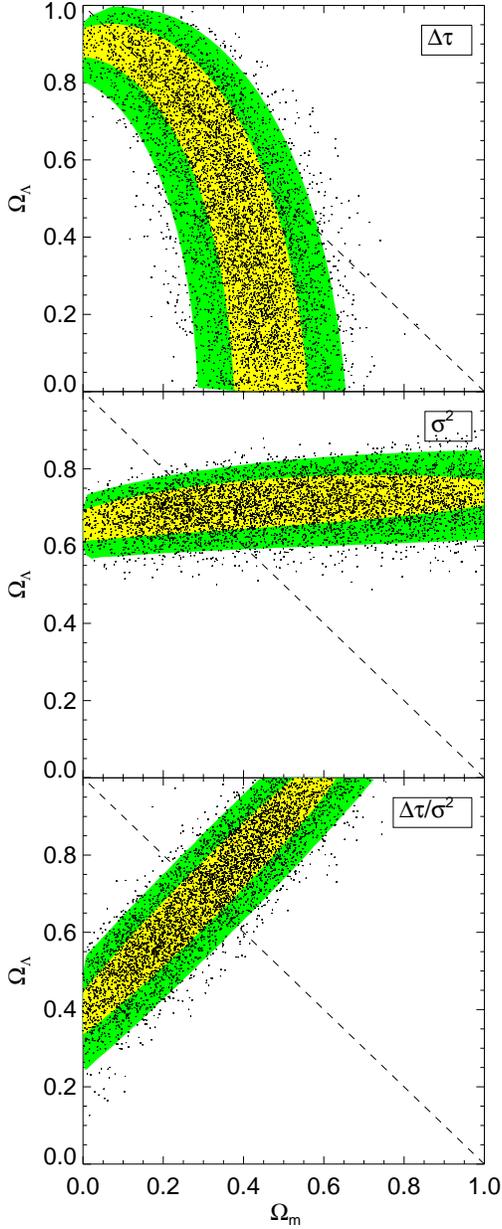}
\caption{MCMC simulations for three different cases: 
$\Delta\tau$ ({\it top, panel}), 
$\sigma^2$ ({\it middle panel}), and $\Delta\tau/\sigma^2$ ({\it bottom panel}). 
125 galaxies were generated with redshifts equally distributed  between 0.1 and 1.1,
source redshifts
between 1.5 and 3.5, and velocity dispersions varying from 100 to 300 \rm km \rm s$^{-1}$. 
For each lens we calculate the Einstein angle for a SIS in a cosmology with 
$\Omega_{\rm m} = 0.3$, $\Omega_\Lambda = 0.7$. $H_0$ was fixed at 70 km s$^{-1}$ Mpc$^{-1}$. {\it Yellow} regions are $1 \sigma$, {\it green} are $2 \sigma$ confidence levels.}
\end{figure}

\subsection{Comparison of methods}
We first illustrate the constraints set in the
$\Omega_{\rm m}$--$\Omega_\Lambda$ plane for 125 simulated  galaxies. Their redshifts
are equally distributed  between 0.1 and 1.1, the source redshifts
between 1.5 and 3.5 and the velocity dispersions in the range 100 to 300 \rm km
\rm s$^{-1}$. For each lens we calculate the Einstein angle from 
 Eq.~3 assuming $\Omega_{\rm m} = 0.3$, $\Omega_\Lambda = 0.7$ and 
$H_0=70$ km s$^{-1}$ Mpc$^{-1}$. 
We constrain the sample to easily detectable systems, thus we include in the 
simulations only 
lenses with $\theta_E$ larger than 0.5\arcsec\ 
\citep{Grillo:2008}. 
We assume simulated 5\% uncertainties in the Einstein angle and run
10000 minimizations of $\chi^2$ for three different observables: 
$\Delta\tau$, $\sigma^2$, and $\Delta\tau/\sigma^2$ (see Fig.~2).

The constraints on the parameter space from the time delay (Fig.~2, top panel) have a non-linear, curved shape 
\citep{Coe:2009},
elongated roughly in  the $\Omega_{\rm m}+\Omega_\Lambda=1$ direction.
The velocity dispersion (Fig.~2, middle panel) gives a good constraint on $\Omega_\Lambda$, but a weak one on $\Omega_{\rm m}$.   Joint observations of time delay and velocity dispersion (Fig.~2, bottom panel) give constraints approximately perpendicular 
to the flat Universe line similarly to Type Ia supernovae. As expected, the proposed cosmic ruler (Eq.~4) gives tighter constraints on both cosmological parameters than the other two methods.
\subsection{One high-redshift lens}
To study the constraints on cosmological parameters from a single lens
we run 10000 minimizations on one system
with parameters similar to an existing lens, MG~2016+112 \citep{Lawrence:1984}. We set the lens redshift to 1.0, the source redshift to 3.27 and the Einstein angle to
$1.7\arcsec$. The velocity dispersion is calculated from   Eq.~3 assuming $\Omega_{\rm m} = 0.3$,
$\Omega_\Lambda = 0.7$. 
We perform simulations for two uncertainty scales (5\%
and 10\%). 
Because the results do not depend on the parameter from which the
uncertainty comes, the simulated error can be understood as the uncertainty in either the
velocity dispersion squared, the time delay, the Einstein angle or a combination of these. 
We present the results from these simulations in Fig. 3 (top row).  The 
probability contours form a wide stripe going across the  parameter space in a
direction 
almost perpendicular to the $\Omega_{\rm m} + \Omega_\Lambda =1$ line. To show the importance of $H_0$ in the simulations we present two cases, first for marginalized $H_0=70 \pm 5 $ km s$^{-1}$ Mpc$^{-1}$ (left column) and second for fixed $H_0=70$ km s$^{-1}$ Mpc$^{-1}$ (right column).
The shift in the marginalized case between the 5\% and 10\% confidence contours is due   to the fact that high values of $H_0$ change the distances calculated from the angular diameter distance equation less than the low ones, thus, to compensate, the region with simulated 10\% uncertainties is shifted towards lower values of  $\Omega_{\rm m}, \Omega_\Lambda$ compared to the 5\% region. 

\begin{figure}[!htbp]
\hspace{-1cm}\includegraphics[scale=0.4]{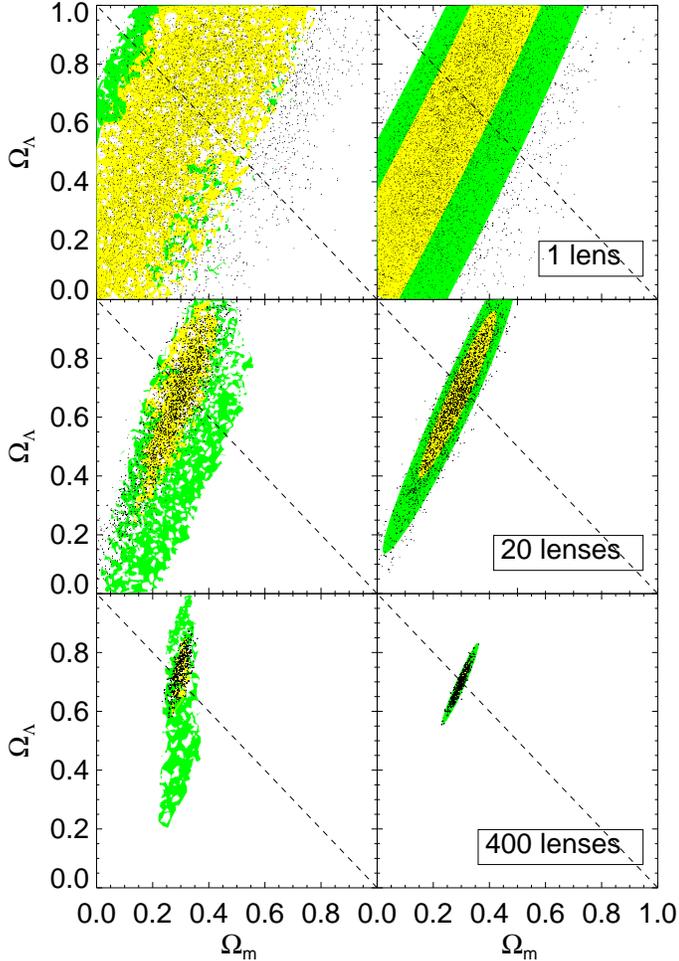}
\caption{MCMC simulations for $\Delta\tau/\sigma^2$. {\it Top row}: One
lens with parameters similar to MG~2016+112 ($z_{\rm L} = 1.0$, $z_{\rm s} = 3.27$, $\theta_{\rm E} =  1.7\arcsec$). 
 {\it Middle row}: 10 lensing systems 
with lens redshifts equally distributed between 0.5 and 1.5. {\it Bottom row}: 400 lensing systems with the same redshifts distribution as for the middle row. 10000
MCMC minimizations were performed.  All three simulations were performed for two uncertainty scales: 5\% ({\it yellow}) and 10\% ({\it green}).  {\it Left column}: The plots marginalized over $H_0$. {\it Right column}: $H_0$ is fixed to 70 km s$^{-1}$ Mpc$^{-1}$.} 
\end{figure}

\subsection{Many lenses}
We also simulate  20 and 400 lensing systems with lens redshifts equally distributed between 0.5 and 1.5 (Fig.~3). 
In this case the
probability of detecting the correct cosmological parameters converges faster around the assumed `real' values $\Omega_{\rm m} = 0.3$, $\Omega_\Lambda = 0.7$,  creating, as a constant probability contour, an ellipse. The simulations show that by observing  20-400 lenses with  small measurement  errors,  cosmological parameters can be well constrained. 
For a 5\% error in the marginalized case we get for 20 lenses:  $\Omega_\Lambda = 0.65_{-0.21}^{+0.17}$, $ \Omega_{\rm m} = 0.30_{-0.07}^{+0.15}$ and for 400 lenses: $\Omega_\Lambda =  0.70_{-0.06}^{+0.05}$, $\Omega_{\rm m} = 0.30_{-0.02}^{+0.02}$
(1 $\sigma$ confidence interval).
\section{Discussion}
In our simulations we have assumed that the lenses have SIS mass distributions.
The SIS profile seems to be a rather good choice, because several studies
based on  dynamics of stars, globular clusters, X-ray halos, etc.\ have shown
that elliptical galaxies have approximately flat circular velocity curves 
\citep{Koopmans:2006b,Oguri:2007,Schechter:2004,Guimaraes:2009,Gerhard:2001}.
The structure of these systems may be considered as approximately
homologous, with the total density distribution (luminous+dark) close to
that of a singular isothermal sphere.
 We stress, however, that the proposed cosmic ruler does not rely on the assumption of
a SIS. For any potential characteristic of the lens galaxies it is sufficient
to assume that the potential (on average) does not change with redshift
(or that any evolution can be quantified and corrected for). In other words,
 $\Delta\tau/\sigma^2 \propto D_{\rm OL}$ for all lens models.
 While we cannot rule out some redshift evolution, it is expected to
be weak \citep{Holden:2009}, given  that lens galaxies are typically massive and hence fairly
relaxed, giving rise to similar structures as their lower-redshift
counterparts.

Nevertheless, like 
for any cosmic ruler, there is a range of possible systematic
uncertainties which must be addressed before it can be used for precision
cosmology. 
Several factors affect the lensing configuration, the velocity dispersion, and the time delays to various degrees:
velocity anisotropy, total mass-profile shape \citep{Schwab:2009} and the detailed density structures (mass
profiles of dark and luminous matter, ellipticities) of the lenses
\citep{Tonry:1983}, mass along the line of sight to the QSO \citep{Lieu:2008}	
and in the environment of the lenses, such as groups and clusters \citep{Metcalf:2005}
(the mass-sheet degeneracy
\citep{Saha:2000,Falco:1985,Williams:2000,Oguri:2007}) and
substructure \citep{Dalal:2002,Maccio:2006,Maccio:2006a,Xu:2009}.
These factors are of the order of our assumed measurement uncertainties.

Fortunately, the velocity anisotropy appears to be small in lens galaxies
\citep{Koopmans:2009,Ven:2003} and in relaxed galaxies in general
\citep{Hansen:2006,An:2006}. The density structures of elliptical galaxies
appear to be fairly universal \citep{Bak:2000,Alam:2002}.
The effects of extrinsic mass are important but usually limited
\citep{Holder:2003} and can be spotted from
anomalous flux ratios or lensing configurations.

To date, there are around 200 known strong gravitational lens 
systems, but only 20 of them have measured time delays, with typical errors 
of 5--10\%. Similar errors are obtained for measured velocity dispersions. 
In other words, applying the method to current data will not provide 
sufficiently tight constraints on $\Omega_{\rm m}$, $\Omega_\Lambda$ to be 
competitive with current cosmological methods. 
 Ultimately therefore, to address systematic errors and to turn
the proposed cosmic ruler into a useful cosmological tool,
a large sample of homogenous systems is needed,
e.g. high-redshift early-type galaxies with $\sigma$, $\Delta\tau$, and the
locations of the images measured with high precision.  From the sample, one 
may eliminate problematic systems, such as systems with a lot of external 
shear, or non-simple lenses.  Moreover,
each lens would have to be modeled in detail, i.e. 
accounting for systematic uncertainties by measuring them 
directly and including the effects in the analysis. 

Future facilities will allow such an experiment.
LSST, SKA, {\em JDEM}, {\em Euclid}, and {\em OMEGA}
\citep{Dobke:2009,Ivezic:2007,Carilli:2004,Marshall:2005,Moustakas:2008}Â
will both find large numbers of new lenses and have the potential to
accurately constrain the time delays. 
And with the   James Webb Space Telescope
({\it JWST}) or future 30--40-m class telescopes    (e.g. TMT -- Thirty Meter Telescope  and E-ELT -- European Extremely Large Telescope), 
velocity dispersions can be obtained to the precision required and
velocity anisotropy can be effectively constrained through integral
field spectroscopy \citep{Barnabe:2009}. This must be coupled with 
high-resolution imaging which can constrain both the density
structures of the lenses and significant mass structures affecting the
lensing. Extrinsic mass and density structures can also be constrained
indirectly by modeling the lensing configuration, including flux ratios
and positions of the images.

The Dark Cosmology Centre is funded by the DNRF.

\end{document}